
\magnification=1200
\hyphenpenalty=2000
\tolerance=10000
\hsize 14.5truecm
\hoffset 1.truecm
\openup 5pt
\baselineskip=24truept
\font\titl=cmbx12
\def\ref{\par\noindent\hangindent 20pt}

\def\mincir{\raise -2.truept\hbox{\rlap{\hbox{$\sim$}}\raise5.truept
\hbox{$<$}\ }}
\def\magcir{\raise -4.truept\hbox{\rlap{\hbox{$\sim$}}\raise5.truept
\hbox{$>$}\ }}
\def\rho{\varrho}
\def\Mdot{\dot M}
\def\etal{{\it et al.\/} }
\def\Menv{M_{env}}
\def\Msyr{M_\odot /yr}
\null
\vskip 1.2truecm

\centerline{\titl ON THE NATURE OF PHOTOSPHERIC}
\centerline{\titl  OSCILLATIONS IN STRONG X--RAY BURSTS}
\vskip 1.5truecm
\centerline{Iosif Lapidus $^{1,3}$, Luciano Nobili $^2$ and Roberto Turolla
$^2$}
\bigskip
\centerline{$^1$ Institute of Astronomy, University of Cambridge}
\centerline{Madingley Road, Cambridge CB3 0HA, UK}
\medskip
\centerline{$^2$ Department of Physics, University of Padova}
\centerline{Via Marzolo 8, 35131 Padova, Italy}
\medskip
\centerline{$^3$ The Royal Astronomical Society Sir Norman Lockyer Fellow}
\bigskip\bigskip\bigskip

\beginsection ABSTRACT

A possible sound origin for the photospheric oscillations in
the X--ray bursting sources 1608-522 and 2127+119 is suggested. It is shown
that standing sound waves in an expanding spherical envelope can have
periods very close to the observed ones. The quite large ratio, $\sim$ 10, of
the periods in the two sources is explained in terms of different wave regimes.
The relevance of sound oscillations to the observed QPO in type II
bursts of the Rapid Burster
is also discussed.

\bigskip\bigskip
\noindent
{\it Subject headings:\/}  stars: individual (1608-522, 2127+119, the Rapid
Burster 1730-335) \ -- \ stars: neutron \ -- \ stars: oscillations \ -- \
X--rays: bursts
\bigskip
\centerline{\it Accepted for publication in ApJ Letters\/}

\vfill\eject

\beginsection 1. INTRODUCTION

Photospheric expansion is commonly observed during very powerful X--ray bursts
and is widely believed to be associated with a supersonic outflow, driven by
the super--Eddington flux produced in the thermonuclear He--burning at
the base of the
envelope (see e.g. Lewin, Van Paradijs \& Taam 1993 for a review on
both theoretical and observational aspects of X--ray bursters).
In a very recent paper (Nobili, Turolla \& Lapidus 1994, hereafter paper I),
we presented a more sophisticated model for
stationary, radiatively driven winds from X--ray bursting neutron star in which
both nuclear burnings at the star surface
and Compton heating--cooling in the outflow were self--consistently treated.
We proposed that the quasi--stationary phase, during
which $L\sim L_{Edd}$, may be regarded as a sequence of steady
wind solutions with decreasing $\Mdot$ which ends when the minimum permitted
value of the mass loss rate for the existence of a stationary, supersonic wind
is reached. The
comparison of our sequence of models with observational data
allows for an estimate of both the hardening factor and the mass accretion
rate onto the neutron star. This approach has been applied to all sources
for which data were available and gave values of $\Mdot_{acc}\magcir 10^{-9}
\Msyr$, in agreement with expectations (Lapidus, Nobili \& Turolla 1994,
hereafter paper II).

Additional information on the physics of bursters' envelopes during
the expansion phase
can be extracted from the second--scale photospheric oscillations registered
so far in two sources. In both cases the light curves showed a quite long,
$\sim$ tens of seconds, flat top, during which the flux was nearly
constant at its maximum level (assumed to coincide with the Eddington
limit); oscillations were seen during
this phase only. Hakucho observations of 1608-522 have detected a
0.65 s oscillation during a $\sim 12$ s flat top (Murakami \etal 1987)
and Ginga data of 2127+119 show a series of oscillations with a
characteristic time scale $\sim 7$ s during a photospheric expansion phase
lasting $\sim 30$ s (Dotani \etal 1990b; Van Paradijs \etal 1990).
In one of earlier theoretical analysis of such phenomena, the evolution of a
gas cloud impinging on the neutron star was studied (Starrfield \etal 1982).
The infall of the cloud resulted in a sharp burst lasting $\sim 1$ s,
followed by a phase of oscillations with a period of 0.2 s. Even though
this period is close to that observed in 1608-522, all the other
characteristics of
the event were not consistent with observations. A time--dependent
study of a largely simplified model of radiatively driven winds
(Yahel \etal 1984) predicted a typical period for photospheric
oscillations of a few tens of milliseconds, and similar results were
obtained by McDermott \& Taam (1987), who examined the non--radial g--modes in
bursting neutron stars atmospheres.
A quite sophisticated theory of oscillations was proposed, more recently,
by Shibazaki \& Ebisuzaki (1989). Although their model successfully reproduced
the 0.65 s oscillation seen in 1608-522,
it is based on a newtonian hydrostatic picture, while it is widely
accepted that a supersonic envelope expansion occurs in very strong bursts.

In this letter we propose that the mechanism responsible for the second--scale
oscillations observed in 1608-522 and 2127+119 is the propagation of
sound waves. The periods of such waves, computed using the
expanding envelope models of paper I, give a correct estimate of the
oscillations timescales in both sources. The present model allows also
for a simple physical explanation for the different, $\sim \, $ factor of
ten,
values of the two observed periods. The relevance of photospheric sound
oscillations in connection with QPO in type II bursts from the Rapid Burster
is also discussed.

\beginsection 2. SOUND OSCILLATIONS IN TYPE I X--RAY BURSTS

Generally speaking, two main types of sound waves exist in a spherically
symmetric medium: radially running waves and standing waves (see e.g.
Landau \& Lifshitz 1987); the latter form when there is a
boundary that can reflect the initial, outward directed
wave. In the expanding neutron star envelope, such a role can be played both
by the sonic surface and by the photosphere, although the physics is different
in the two cases. Below the sonic radius $R_s$ the flow is
subsonic and a sound wave can propagate both upstream and downstream,
while only a strongly attenuated part of the wave survives
at $R > R_s$. In this sense, the initial outgoing disturbance will be mostly
reflected backwards near $R_s$, giving rise to a standing sound wave. What
makes the photosphere a peculiar surface, as far as wave propagation is
concerned, is the fact that the sound speed drops drastically across it.
Below the photosphere, in fact, LTE holds and matter and radiation are strongly
coupled, forming a single fluid with pressure $P=P_{gas} + P_{rad} = P_{gas}
(1+\alpha )$. On the contrary, outside the photosphere the
radiation is no longer
in equilibrium with the gas, no radiative contribution to the total pressure
is present, although photons can still exchange momentum with matter, and
$P=P_{gas}$.
The transition is gradual and it occurs around the radius at which the
effective
optical depth is about unity. In our model envelopes, the width of the
transition layer is $\Delta R\sim R$, so that the approximation of a
discontinuous
transition may be used only to provide order of magnitude estimates. Clearly
temperature and gas density are continuous across the surface, while,
because of the change in the equation of state, pressure,
energy density and sound speed are not. In particular, the ratio of sound
speeds below (region 1) and above (region 2) the photosphere is
\vfill\eject
$${{v_s^{(1)}}\over{v_s^{(2)}}} = \left[1+{{32\alpha^2}\over{5(1+8\alpha )}}
\right]^{1/2}\, ,$$
which implies a reflection coefficient for acoustic waves (see again
Landau \& Lifshitz)

$${\bf R}\sim \left({{v_s^{(1)}/v_s^{(2)} - 1}\over{v_s^{(1)}/v_s^{(2)} + 1}}
\right)^2\, .\eqno(1)$$
Here we denote by $R_{ph}$ the lower border of the transition layer, i.e.
the radius at which LTE begins to break, and where the ratio of the radiation
energy density to the blackbody one, $u/aT^4$, starts to
drop below unity. The
sound speed at $R_{ph}$ is $v_s(R_{ph}) = v_s^{(1)}\propto \sqrt{\alpha T}$
(in all our models $\alpha(R_{ph})\gg 1$) while, at larger radii,
$v_s(R) = v_s^{(2)}\propto \sqrt{T}$.

Before examining in some more details the different wave propagation modes, we
present some parameters of the wind models, calculated in
paper I, which are relevant to our present analysis.
Results are summarized in table 1 where,
for each $\Mdot$, the following quantities are listed:
the sonic radius, $R_s$, the ``sonic'' sound crossing time $R_s/v_s(R_s)$,
the photospheric radius $R_{ph}$, computed assuming that  $u/aT^4= 0.9$,
$\alpha(R_{ph})$, and the corresponding ``photospheric'' sound crossing time,
$R_{ph}/v_s(R_{ph})$; all the data refer to a
neutron star of mass $M_* = 1.5\, M_\odot$, radius $R_* = 13.5$ km and to an
envelope with nearly solar chemical composition.
As can be seen from the table, the photosphere is always inside the sonic
radius for wind models with $\Mdot < 50\, \Mdot_{Edd}$ and,
in both 1608-522 and 2127+119, the mass loss rate at the beginning of the
quasi--stationary wind phase was estimated to be below this value (see
paper II). We mention that the photospheric radii as defined here exceed
slightly ($\mincir 10\%$) those of paper I, where the photosphere was taken
to be at $\tau_{eff} = 3$.

Let us consider now the possible regimes of sound oscillations.
For a standing wave to be formed, the initial spherical diverging
wave must reach the photosphere, reflect back and return
to the base of the envelope, so it will take, at least, a time
$t_{ph}\sim 2 R_{ph}/v_s^{(1)}$ for this standing wave to get stabilized.
Data in table 1 show that $t_{ph}$ is shorter than the flat--top durations
and ``photospheric'' standing waves can settle in both sources. The possible
occurrence of the second type of standing waves, the ``sonic'' one, is
restricted to 2127+119, because only in this source $t_{s}\sim 2 R_s/v_s(R_s)$
is shorter than the flat--top duration.
In order to derive a simple estimate of the
pulsational eigenfrequencies, we assume that the local sound speed is constant
over the entire shell $R_{in}<R<R_{out}$ and that $R_{in}$ can be taken to be
zero to all practical purposes. Here $R_{in} = R_*$, $R_{out} =R_{ph}$
(for ``photospheric'' waves) and $R_{out} = R_s$ (for ``sonic'' waves).
Both these hypothesis are reasonable since
numerical wind solutions show that $v_s$ does not vary by orders of magnitude
and $R_*\ll \, R_{out}$ (see also Table 1).
The velocity potential for a monochromatic standing spherical wave has the
form (Landau \& Lifshitz)

$$\varphi=A\, {\sin{kr}\over r} e^{-i \omega t} \, \eqno(2)$$
and from this expression the eigenfrequencies equation can be easily obtained
once a boundary condition at $r=R_{out}$ is specified.
The fixed boundary condition, $v=\partial \varphi/\partial r =0$ at
$r=R_{out}$, results in

$$\tan z\, = \, z \,, \eqno(3)$$
where $z=\omega R_{out}/v_s$. The corresponding periods are

$$P_n^{fix} \, \sim {4\over {1+2n}}\cdot {R_{out}\over v_s}\,, \,\,
\,\,\,n=1,2,...
\eqno(4)$$
In the case of a free oscillating boundary,
$\delta p \,=\,-\varrho \partial \varphi/\partial t\, =\, 0$
at $r=R_{out}$, the eigenfrequencies equation becomes
$$ \sin z\,=\,0\,,\eqno(5)$$
and the periods of the eigenmodes are

$$P_n^{free} \,=\, {2\over n}\cdot {R_{out}\over v_s}\,, \,\,
\,\,\,n=1,2,...
\eqno(6)$$

For the burst in 1608-522 observed with Hakucho
the mass loss rate at the beginning of the wind phase, i.e. when the
luminosity has just reached the Eddington level, was estimated to be
$\sim 26 \dot M_{Edd}$ (paper II). From the table it can be seen that
the sound crossing time ranges from $\sim 0.4$ to $\sim 0.1$ s, as the mass
of the envelope decreases during the wind phase.
The free boundary condition at the photosphere seems to be more
appropriate than the fixed one, since there is no physical mechanism to keep
the photosphere fixed in space. The period of the principal mode
is $P_1\sim 2R_{ph}/v_s(R_{ph})\sim 0.8 - 0.1$ s, which is a fairly good
estimate for the $\sim 0.65$ s oscillations observed in 1608-522.
Although not only the principal mode may be present,
the amplitudes of higher order
oscillations are much more reduced by damping: in fact, denoting by
$\gamma_n$ the $n$--th mode sound absorption coefficient, it is
$\gamma_n/\gamma_1 = \omega_n^2/\omega_1^2 = n^2$.

In 2127+119 photospheric oscillations with a characteristic time scale
$\sim 7$ s were observed with Ginga during a $\sim 30$ s flat top of the
light curve. In paper II, the initial envelope mass was estimated to be
$\Menv \sim 8\times 10^{22}$ g, corresponding to $\Mdot\sim 16\,
\Mdot_{Edd}$. Now the flat top duration is longer than $t_s$, so it is
reasonable to expect that also a ``sonic'' standing wave has time to develop
because, for a typical value of $\alpha \sim 800$, about 15 \%
of the initial disturbance is transmitted through the photosphere (see
eq. [1]) and will form later the ``sonic'' standing wave. We note that now
both the free and the
fixed boundary conditions provide radial oscillations of the photosphere,
since, for the fixed one, the node is at $R_s$ and not at $R_{ph}$. Equations
(4) and (6) show, however, that the differences in the eigenfrequencies are
relatively small, so, despite our ignorance about the exact form of the
boundary condition at $R_s$, we can get a fair estimate of the period.
Referring to the free--boundary spectrum, the fundamental mode has
$P_1\sim 22$ s and can hardly be detected, being too close to
the total duration of the flat--top phase. The second harmonic
has a period  $P_2\sim 11$ s which is close to the observed
$\sim 7$ s periodicity; of course this mode is weaker than the fundamental
one, but it should dominate the Fourier spectrum in the absence of the
main one. Although the simultaneous detection of both kinds of standing waves
in 2127+119 cannot be excluded, short wavelength ``photospheric'' modes
are more severely damped with respect to the longer ``sonic'' modes.

Another observational fact which supports the hypothesis of
radial photospheric oscillations is the phase lag of about
$180^{\circ}$ between the oscillations in the low-- and high--energy bands.
The constant luminosity ($L\simeq L_{Edd}$) radial oscillations of the
photosphere
result in simultaneous oscillations of the photospheric temperature
and the degree of Comptonization of radiation crossing the translucent
region, and therefore in the
simultaneous oscillations of the whole spectrum
registered by a distant observer. As the temperature
reaches its maximum, the counts in the hard bands are the maximum, while
those in the soft bands are the minimum, and vice versa. It means that the
strong anticorrelation between the counts at low and high energies
oscillating with the same frequency, provides the additional evidence in
support of our theory of sound waves.

\beginsection 3. QPO IN TYPE II X--RAY BURSTS

The physics of ``abnormal'' type II bursts is still unclear, but the
currently dominating idea is that these bursts are due to some sort of
instability in the inner regions of the accretion flow onto the
neutron star. A detailed review of the observations of the Rapid Buster
(RB) can be found in Lewin \etal
Type II bursts, too, show some observational evidence of photospheric
expansion.
Kunieda \etal (1984) and Tan \etal (1991), in fact, have shown that there is a
correlation between the color temperature, $T_{col}$, and the
luminosity, $L$, observed during the flat tops of relatively long
bursts. It was found that $L \propto T_{col}^6$ and this led Tan
\etal to suggest that the radiation released during a type II burst
comes from a photosphere whose radius is larger for higher accretion
rates. Quasi--periodic oscillations (QPO) with frequencies $\sim 2$ Hz
were discovered in
type II bursts from the RB with Hakucho, and a further analysis of
these QPO indicated that, most probably, they have no relation
to other forms of QPO observed in many bright LMXBs (see e.g. Lewin \etal for
references). An indirect confirmation of this fact was provided by
Rutledge \etal (1993), who have shown that the RB is neither an
atoll nor a Z source in the classification of Hasinger \&
Van der Klis (1989). Lubin \etal (1991) showed that the QPO
centroid frequencies in type II bursts range from $\sim 2$ to
$\sim 7$ Hz and they are strongly anticorrelated with the peak flux in
the burst. It was proposed by Dotani \etal (1990a) that QPO can be
described by changes in the photospheric radius and
temperature. Lewin \etal (1991) have given some arguments why this
possibility is more likely than the changes in the temperature of a
blackbody emitter with constant apparent area.

Here we suggest that QPO in type II bursts could have the same origin as
photospheric oscillations in ``normal'', type I bursts. We emphasize,
however, that no quasi--stationary, Eddington luminosity phase is
attained in type II bursts, so our
wind models are not directly applicable for describing
the expansion/contraction phase of
the Rapid Burster. Still, it would be reasonable to assume that, if the
luminosity is sub--Eddington or super--Eddington during a short unstationary
stage, the structure of the outflow is similar to that of our models
representing the final part of the expansion/contraction phase, i.e.
those ones in the lower end of the $\Menv$ (or $\Mdot$) range.
The characteristics of these models may provide order of magnitude
estimates for various quantities in the expanded RB envelope.
The frequency of the principal mode of the ``photospheric'' standing wave
is $\approx 10$ Hz for the models with the lowest possible
$\dot M$ (last lines in Table 1), and  $\approx 2$ Hz for
an envelope with $\Mdot \sim 20 \dot M_E$, as that one of both 1608-522
and 2127+119 at the beginning of the expansion/contraction phase.
We see that the $\sim$ 2-7 Hz range of the QPO centroid frequencies of the RB
may be easily reproduced in frame of our hypothesis. Let us consider now how
the sound frequency correlates with the peak flux in the burst.
In our wind solutions the larger is the envelope mass, the larger is
the photospheric radius and the lower is the temperature
at the photosphere. As it may be seen from the last
column of Table 1, the period of the ``photospheric'' standing waves
monotonically decreases with decreasing $\Mdot$.
The sound oscillations frequency, therefore,
will be higher for the less energetic bursts and
this is indeed consistent with the observed anticorrelation of the QPO
frequency with the average burst peak flux.
It is also not surprising that, being of sound origin,
the QPO in the Rapid Burster are qualitatively different
from the standard QPO seen in many bright LMXBs which are
probably due to dynamical effects.

\beginsection 4. CONCLUSIONS

We have presented a simple model for the photospheric oscillations observed
during type I X--ray bursts from 1608-522 and 2127+119, and in
type II bursts of the Rapid Burster. In particular,
we have shown that the periods of standing sound waves in the expanding
envelopes analyzed in paper I are very close to the observed ones.
The
quite large, about a factor of 10, difference in the observed periods is
explained in terms of the different wave regimes in the two sources:
``photospheric'' and ``sonic'' standing sound waves.
The acoustic wave hypothesis could also explain the
QPO in type II bursts from the Rapid Burster; both the QPO
centroid frequencies and the strong anticorrelation of the
centroid frequency with the peak flux in the burst are, in fact,
consistent with the ``photospheric'' standing sound wave scenario.

The existence of two different wave regimes can lead to significant
observational predictions.
A number of bursters exhibits, in fact, strong bursts with
precursors which have long expansion phases.
The ``sonic'' standing wave has, therefore, enough
time to form and oscillations with periods up to
$\sim 25$ s may be revealed in the light curves flat tops.
Both the re--examination of archive data and
new observations with
X--ray telescopes on board satellites of the current generation, like ASCA,
may definitely access the existence of such oscillations.

\beginsection ACKNOWLEDGMENTS

We are indebted to an anonymous referee whose constructive criticism
helped in improving an earlier version of this paper.

\beginsection REFERENCES

\ref{Dotani, T., Mitsuda, K., Inoue, H., Tanaka, Y., Kawai, N., Tawara, Y.,
Makishima, K., Van Paradijs, J., Penninx, W., Van der Klis, M., Tan, J.,
\& Lewin, W.H.G. 1990a, ApJ, 350, 395}
\ref{Dotani, T., Inoue, H., Murakami, T., Nagase, F., Tanaka, Y., Tsuru,
T., Makishima, K., Ohashi, T., \& Corbet, R.H.D. 1990b, Nature, 347, 534}
\ref{Hasinger, G., \& Van der Klis, M. 1989, A\&A, 225, 79}
\ref{Kunieda, H., Tawara, Y., Hayakawa, S., Masai, K., Nagase, F., Inoue,
H., Koyama, K., Makino, F., Makishima, K., Matsuoka, M., Murakami, T., Oda,
M., Ogawara, Y., Ohashi, T., Shibazaki, N., Tanaka, Y., Kondo, I.,
Miyamoto, S., Tsunemi, H., \& Yamashita, K. 1984, PASJ, 36, 215}
\ref{Landau, L.D., \& Lifshitz, E.M. 1987, Fluid Mechanics, (Oxford: Pergamon)}
\ref{Lapidus, I., Nobili, L., \& Turolla, R. 1994, ApJ, in press (Paper II)}
\ref{Lewin, W.H.G., Lubin, L.M., Van Paradijs, J., \& Van der Klis, M. 1991,
A\&A, 248, 538}
\ref{Lewin, W.H.G., Van Paradijs, J., \& Taam, R.E. 1993, Space Sci.
Rev., 62, 223}
\ref{Lubin, L.M., Stella, L., Lewin, W.H.G., Tan, J., Van Paradijs, J.,
Van der Klis, M., \& Penninx, W. 1991, MNRAS, 249, 300}
\ref{McDermott, P.N., \& Taam, R.E. 1987, ApJ, 318, 278}
\ref{Murakami, T., Inoue, H., Makishima, K., \& Hoshi, R. 1987, PASJ, 39,
879}
\ref{Nobili, L., Turolla, R., \& Lapidus, I. 1994, ApJ, in press (Paper I)}
\ref{Rutledge, R.E., Lubin, L.M., Lewin, W.H.G., Van Paradijs, J.,
\& Van der Klis, M. 1993, MNRAS, submitted}
\ref{Shibazaki, N., \& Ebisuzaki, T. 1989, PASJ, 41, 641}
\ref{Starrfield, S., Kenyon, S., Sparks, W.M., \& Truran, J.W. 1982, ApJ,
258, 683}
\ref{Tan, J., Lewin, W.H.G., Lubin, L.M., Van Paradijs, J., Penninx, W.,
Damen, E., \& Stella, L. 1991, MNRAS, 251, 1}
\ref{Van Paradijs, J., Dotani, T., Tanaka, Y., \& Tsuru, T. 1990, PASJ, 42,
633}
\ref{Yahel, R.Z., Brinkmann, W., \& Braun, A. 1984, A\&A, 139, 359}

\vfill\eject

%
\null
\vskip 1.5truecm
\centerline{Table 1}\medskip
\centerline{Parameters of Model Atmospheres}
$$\vbox{\tabskip=1em plus2em minus.5em
\halign to\hsize{\hfil # \hfil & \hfil # \hfil & \hfil # \hfil &
 \hfil # \hfil  & \hfil # \hfil & \hfil # \hfil &
\hfil # \hfil & \hfil # \hfil &\hfil # \hfil &\hfil # \hfil\cr
$\Mdot $ & $R_s$ & $R_s/v_s(R_s)$ & $R_{ph}$ & $\alpha(R_{ph})$
& $R_{ph}/v_s(R_{ph})$ & & & & \cr
$(\Mdot_{Edd})$  & $(10^3 {\rm km})$ & $ ({\rm s}) $
& $ (10^3 {\rm km})$ & $ (10^2) $  & ({\rm s}) & &  & & \cr
%
%
\noalign{\medskip}

50.2 & 5.9 & 27.9 & 2.20 & 4.6 & 1.47 & & & &\cr
26.7 & 4.0 & 14.9 & 0.96 & 6.6 & 0.39 & & & &\cr
19.8 & 3.3 & 10.9 & 0.65 & 7.8 & 0.21 & & & &\cr
12.7 & 3.9 & 11.4 & 0.49 & 10.3& 0.10 & & & &\cr
8.8 & 5.0 & 13.4 & 0.27 & 11.2& 0.06 & & & &\cr
6.7 & 6.0 & 15.2 & 0.23 & 13.3& 0.05 & & & &\cr
5.9 & 7.0 & 17.3 & 0.21 & 13.8& 0.04 & & & &\cr
4.8 & 9.5 & 23.5 & 0.19 & 13.9& 0.04 & & & &\cr
2.8 & 16.4 & 38.0 & 0.16 & 14.8& 0.03 & & & &\cr
 }}$$

\vfill\eject
\bye